\documentclass[12pt]{article}
\usepackage{epsfig}

\def\beq{\begin{equation}}
\def\eeq{\end{equation}}
\def\beqn{\begin{eqnarray}}
\def\eeqn{\end{eqnarray}}

\newcommand{\ccaption}[2]{
\begin{center}
\parbox{0.85\textwidth}{
\caption[#1]{\small{\it{#2}}}
}
\end{center}
}

\newcommand\sss{\scriptscriptstyle}
\newcommand\mug{\mu_\gamma}

\newcommand\muf{\mu_{\sss F}}
\newcommand\mur{\mu_{\sss R}}

\newcommand\as{\alpha_{\sss S}}

\newcommand\qq{{\scriptscriptstyle Q\overline{Q}}}
\def\pt{p_{\scriptscriptstyle T}}

\begin{document}
\noindent {\bf G. Ridolfi}\footnote{INFN Sezione di Genova, Genoa, Italy}
\bigskip

\noindent {\bf PHOTOPRODUCTION OF HEAVY QUARKS AT HERA}
\bigskip

\noindent
{\it Talk given at the 3$^{rd}$ UK Phenomenology Workshop on
HERA Physics, Durham (UK), September 20$^{th}$-25$^{th}$, 1998.}
\bigskip

Heavy-flavoured hadrons are mainly produced at HERA in those events in
which the virtuality of the photon exchanged between the electron and
the proton is very small (see R. Thorne's talk~\cite{Thorne} for a
review on heavy quark production in deep-inelastic scattering.) The
electron, in this case, is equivalent to a beam of on-shell photons
carrying a fraction of the electron energy distributed according to
the Weizs\"acker-Williams function~\cite{WW}. The underlying
production mechanism is therefore a photoproduction one, which has
been studied extensively in fixed-target experiments (see
ref.~\cite{FMNRbook} for a complete bibliography); HERA offers the
opportunity of testing the predictions of perturbative QCD for this
process in an energy regime which is totally unexplored, the available
centre-of-mass energy being about one order of magnitude larger than
in fixed-target experiments. This is an interesting opportunity, since
photoproduction is much less affected by nonperturbative and higher
order effects than hadroproduction.

A well-known feature of photoproduction processes is that the incoming
photon may fluctuate into a hadronic state before undergoing a hard
collision. The corresponding contribution to the cross section is
referred to as {\it hadronic} (or {\it resolved}), to contrast with
the case when the photon directly interacts with the hadron ({\it point-like}
or {\it direct}.)  Therefore, a differential photon--hadron cross
section can be written as the sum of a point-like and a hadronic
photon contribution:
\beq
d\sigma^{(\gamma {\sss H})}(P_\gamma,P_{\sss H})=
d\sigma^{(\gamma {\sss H})}_{\rm point}(P_\gamma,P_{\sss H})
+d\sigma^{(\gamma {\sss H})}_{\rm hadr}(P_\gamma,P_{\sss H})\,.
\label{fullxsec}
\eeq
According to factorization theorems, the two contributions can be
expressed as 
\beqn
d\sigma_{\rm point}^{(\gamma {\sss H})}(P_\gamma,P_{\sss H})&=&\sum_j\int dx
f^{({\sss H})}_j(x,\muf)
d\hat{\sigma}_{\gamma j}(P_\gamma,xP_{\sss H},\as(\mur),\mur,\muf,\mug),
\label{pointcomp}
\\
d\sigma_{\rm hadr}^{(\gamma {\sss H})}(P_\gamma,P_{\sss H})
&=&\sum_{ij}\int dx dy
f^{(\gamma)}_i(x,\mug) f^{({\sss H})}_j(y,\muf^\prime)
\nonumber \\*&&\phantom{\sum_{ij}\int dx}\times
d\hat{\sigma}_{ij}(xP_\gamma,yP_{\sss H},
\as(\mur^\prime),\mur^\prime,\muf^\prime,\mug)\,,
\label{hadrcomp}
\eeqn
where $f^{(\gamma)}_i$ and $f^{({\sss H})}_j$ are the parton densities
in the photon and in the hadron respectively. The point-like and the
hadronic components of the photoproduction cross sections are very
closely related, and only their sum is physically meaningful. In fact,
the separation of a cross section into a point-like and a hadronic
component is ambiguous beyond leading order: different choices of the
factorization schemes lead to different definitions of the two
components. This is explicitly shown in eqs.~(\ref{pointcomp})
and (\ref{hadrcomp}): the same arbitrary scale $\mug$, related to the
subtraction of collinear singularities of quarks emitted by the
incoming photon, appears in both equations.

The parton densities in the photon are quite soft. Therefore, the hadronic
component is only important for large CM energies and small masses for
the produced system, which is precisely the case for charm production
at HERA. This process can therefore be used to constrain the
densities in the photon, which are experimentally very poorly known at
present.

\begin{figure}
\centerline{\epsfig{figure=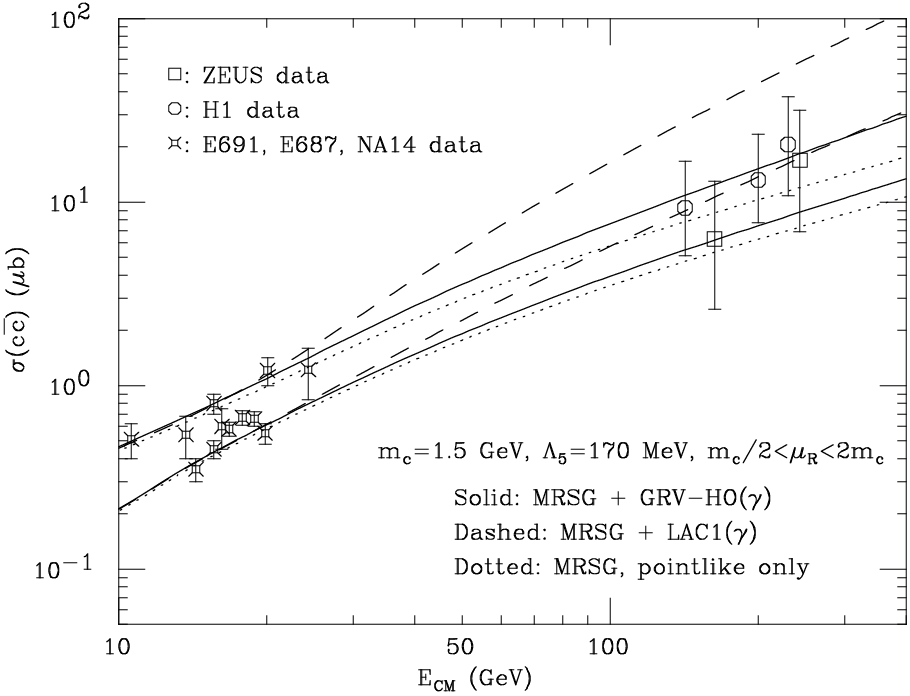,width=0.7\textwidth,clip=}}
\ccaption{}{ \label{f:sig_vs_ecm}
Total cross section for the photoproduction of $c\bar{c}$ pairs,
as a function of the $\gamma p$ centre-of-mass energy:
next-to-leading order QCD predictions versus experimental results.}
\end{figure}

In fig.~\ref{f:sig_vs_ecm} I present measured values of the total charm
cross section in fixed target and HERA experiments, superimposed on
the theoretical predictions, for three different choices of the
parametrization of the photon parton densities. Each pair of curves
represents the change in the theoretical predictions due to variations
of the renormalization and factorization scales; the corresponding
uncertainty is quite large, due to the small value of the charm quark
mass, which sets the energy scale of the process. It is also clear
that the impact of the hadronic component is negligible at fixed
target energies, while it becomes important in the HERA energy range.
The data show a satisfactory agreement with the curves obtained by
choosing $m_c=1.5$~GeV and MRSG~\cite{MRSG} and GRV-HO~\cite{GRVHO}
for the partonic densities in the proton and the photon respectively.
Remarkably, a single choice of the input parameters allows a good
description of the data in the whole energy range considered.

It is apparent from fig.~\ref{f:sig_vs_ecm} that no definite
conclusion can be drawn on the photon densities. This is partly
because of the presence of inconsistencies in the low-energy data, and
partly because of the intrinsic uncertainties of the HERA
measurements. In fact, the experiments are only sensitive to
production in the central rapidity region; furthermore, a
small-$p_{\sss T}$ cut is applied to the data, in order to clearly
separate the signal from the background.  This kinematical region
typically involves small-$x$ values, and the extrapolation is
therefore subject to uncertainties due to our ignorance of the
small-$x$ behaviour of the parton densities. 

Single-inclusive distributions have also been measured by ZEUS and H1;
a detailed description of the data compared to theoretical predictions
is given in the contribution of J.~Cole to these Proceedings~\cite{Cole}.
A general comment to these studies is that the agreement appears to
be satisfactory; however, it is necessary to wait for more statistics
to draw significant conclusions.  In fact, if data with larger
statistics were available, it would be possible to consider the
production processes initiated by very energetic photons only, instead
of integrating over the whole photon spectrum using the
Weizs\"acker-Williams distribution.

At HERA energies, large terms proportional to powers of
$\log(S/m_c^2)$ appear in the charm cross section, that may spoil the
convergence of the perturbative series.  The problem of resumming
these terms ({\it small-x effects}) has been studied in
ref.~\cite{smallx} in different contexts.  In ref.~\cite{FMNRheratot}
it has been estimated that these effects are negligible with respect
to the other sources of uncertainty on the theoretical predictions.

A similar problem arises when the transverse-momentum distribution is
considered; in this case, terms proportional to powers of
$\log(p_{\sss T}/m_c)$ are present in the coefficients of the
perturbative expansion, which are potentially dangerous at large
$\pt$.  These logarithms can be resummed by observing that, at high
$p_{\sss T}$, the heavy-quark mass is negligible, and by using
perturbative fragmentation
functions~\cite{Cacciari96,Kniehl95,MeleNason}.  It has been checked
that the fixed-order and the resummed results of \cite{Cacciari96}
agree in a very wide range in $p_{\sss T}$.  Recently, a formalism has
been developed~\cite{CacciariGrecoNason} that allows matching the two
approaches (the fixed-order calculation with $m\not=0$ and the
resummed result for massless quarks). It has however been applied only
to hadroproduction processes.

Let us now briefly consider $b$ production.  This case is more reliably
described in perturbative QCD, because of the larger value of the
$b$-quark mass.  All the theoretical uncertainties are strongly reduced
with respect to the charm case.  Also, the hadronic component of the
photon is less important in $b$ production, which involves larger values
of $x$.  It should be remarked that a statistically significant study
of bottom production will be possible at HERA only with a luminosity
upgrade (the total rates are about a factor 200 smaller than those of
charm production), and in any case a comparison with theoretical
predictions will only be possible considering electroproduction in the
Weizs\"acker--Williams approximation.

\begin{figure}
\centerline{\epsfig{figure=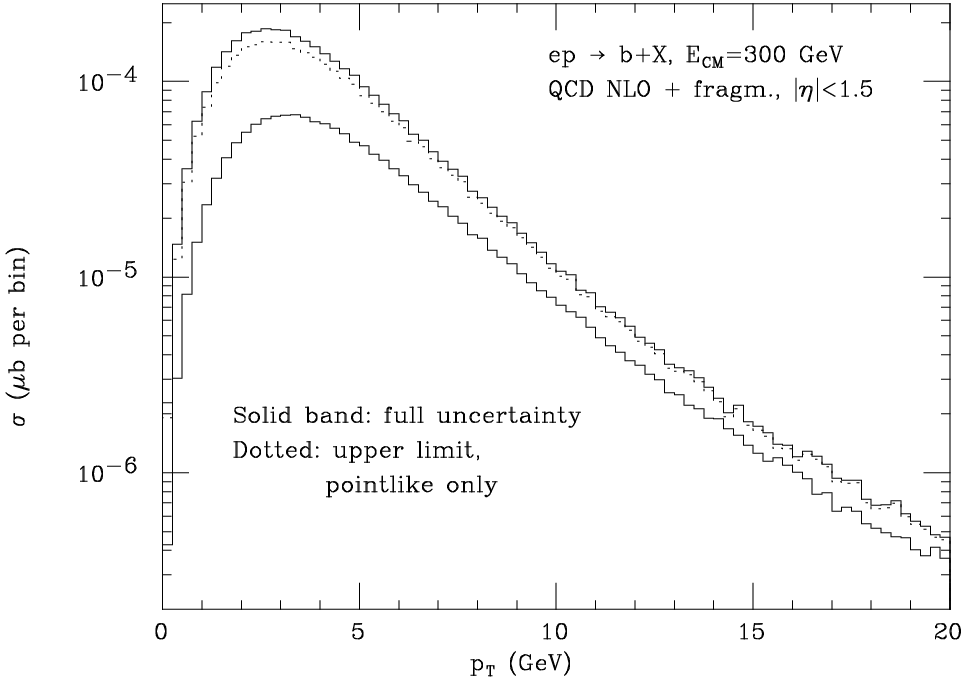,width=0.7\textwidth,clip=}}
\ccaption{}{ \label{f:b_pt_at_HERA}
Full uncertainty on the transverse-momentum distribution for bottom
electroproduction (Weizs\"acker-Williams approximation)
with Peterson fragmentation and a pseudorapidity cut.}
\end{figure}

Figure~\ref{f:b_pt_at_HERA} shows that the transverse momentum
distribution of $b$ quarks at HERA can be predicted by perturbative
QCD quite accurately. Even with the LAC1~\cite{LACone}
set the hadronic component
affects the prediction only marginally; this is basically a
consequence of the applied pseudorapidity cut.

The first measurements of the $b$ production cross section at HERA
have recently been presented by the H1 Collaboration\cite{Tsipolitis}.

If the integrated luminosity of the HERA collider will increase to
100~pb$^{-1}$ or more, it is conceivable to perform analyses of
double-tagged charm events~\cite{Eichler96}, and therefore to study
double-differential cross sections. This would be a precious
opportunity in many respects; for example, a direct determination of
the proton gluon density $f_g^{(p)}$ \cite{glufromc,FMNRglu} would
become possible.

In photon--hadron or electron--hadron collisions, the gluon density
enters the cross section in a simpler way than in the case of hadronic
collisions. If one could determine the invariant mass and rapidity of
the produced system, the leading-order cross section would be directly
proportional to $f_g^{(p)}$ and to other calculable factors. For
example, the LO heavy-flavour electroproduction cross section can be
written, in the Weizs\"acker--Williams approximation, and neglecting
the hadronic component, in the following way:
\beq
\frac{d \sigma^{(0)}}{d y_{\qq} \, dM_{\qq}^2 }=
x_g\frac{d \sigma^{(0)}}{d x_g \, dM_{\qq}^2 }= \frac{1}{E^2}\;
f^{(e)}_{\gamma}(x_\gamma,\mu_0^2)
f^{(p)}_{g}(x_g,\muf^2)
\hat{\sigma}^{(0)}_{\gamma g}(M_{\qq}^2),
\label{ptoqq}
\eeq
where $M_{\qq}$ is the invariant mass of the heavy-quark pair, and
$y_{\qq}$ is the rapidity of the pair in the electron--proton
centre-of-mass frame (positive rapidities are chosen in the direction
of the incoming proton); $E=\sqrt{S}$ is the electron--proton
centre-of-mass energy, and
\beqn
x_\gamma&=&\frac{M_{\qq} }{E} \exp(-y_{\qq}),
\\
x_g&=&\frac{M_{\qq} }{E} \exp(y_{\qq}).
\label{xgdef}
\eeqn
The function $f^{(e)}_{\gamma}$ is the Weizs\"acker--Williams distribution.
All quantities on the right-hand side of eq.~(\ref{ptoqq}) are
calculable, except for $f_g^{(p)}$, which can therefore be measured.
\begin{figure}[htb]
\centerline{\epsfig{figure=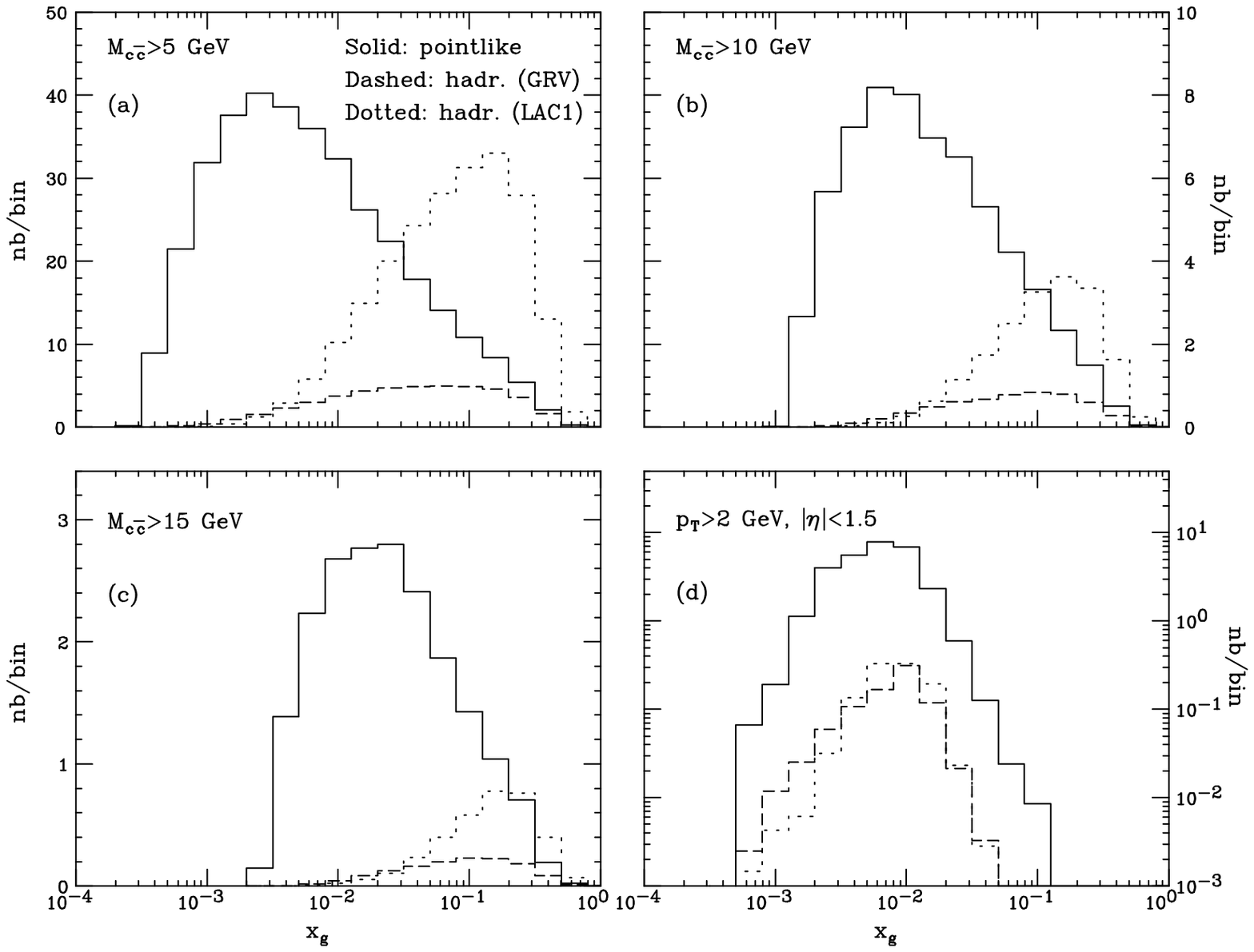,width=0.8\textwidth,clip=}}
\ccaption{}{ \label{f:xgdistr}
$x_g$ distribution in $ep$ collisions (Weizs\"acker-Williams approximation)
at HERA, for $m_c=1.5$~GeV and the MRSG parton densities in the proton.}
\end{figure}
The inclusion of radiative corrections up to NLO is straightforward, as
explained in ref.~\cite{FMNRglu}.

The method outlined above for the extraction of the gluon density in
the proton will only work in those kinematical regions where the
hadronic component is suppressed.  This possibility is illustrated in
fig.~\ref{f:xgdistr}, where the NLO QCD distribution in the variable
$x_g$ in electron-proton collisions at $\sqrt{S}=314$~GeV is
presented.  The partonic densities in the proton are given by the MRSG
set, while both the LAC1 and GRV-HO sets for the photon are considered.
Figures~\ref{f:xgdistr}a)--\ref{f:xgdistr}c) show the effect of
applying a cut on the invariant mass of the pair. Even in the case of
the smallest invariant-mass cut, there is a region of small $x_g$
where the hadronic component is actually negligible.  Increasing the
invariant-mass cut, the hadronic component decreases faster than the
point-like one.  This is because for large invariant masses the
production process of the hadronic component is suppressed by the
small value of the gluon density of the photon at large $x$. By
pushing the invariant-mass cut up to 20 GeV, it turns out that the
point-like component dominates over the hadronic one for $x_g$ values
as large as $10^{-1}$. Therefore, it can be concluded that the
theoretical uncertainties affecting the charm cross section, in the
range of $10^{-3}<x_g<10^{-1}$, are small enough to allow for a
determination of the gluon density in the proton by using
invariant-mass cuts to suppress the hadronic component.
Figure~\ref{f:xgdistr}d) shows the effect, on the $x_g$ distribution,
of a small-$p_{\sss T}$ and a pseudorapidity cut applied to both the
charm and the anticharm, as usually done in HERA analyses. It is
interesting to note that in this case the point-like component is
dominant in the whole kinematically accessible range, even without
applying an invariant-mass cut.

Recently, a determination of the gluon density based on single-inclusive charm
measurements has been performed by the H1
Collaboration~\cite{Honeglu}. The results are encouraging: the 
gluon density $g(x,\mu^2)$ has been determined in the range
$7.5\cdot 10^{-4}<x<4\cdot 10^{-2}$, at average scales  $\mu^2=25$
to $50$ GeV$^2$, and the result is in good agreement with 
those obtained from the analysis of scaling violations of $F_2$.

The possibility of performing experiments at HERA with longitudinally
polarized beams has also been considered \cite{DeRoeck97}.  At
leading order in QCD, the heavy-flavour production cross section in
polarized $ep$ collisions is proportional to the polarized gluon
density in the proton, $\Delta g$.  Therefore, data on charm
production could in principle be used to measure $\Delta g$ directly,
as in the unpolarized case.  This possibility was first suggested in
refs.~\cite{deltag} and reconsidered in ref.~\cite{FRdeltag} and in
ref.~\cite{SVdeltag}; in this last paper the impact of the hadronic
component has been estimated.  Finally, a full NLO computation of
heavy quark photoproduction with polarized beams has been presented
in ref.~\cite{Bojak}.

The result is that total cross section asymmetries for the point-like
component are quite small in absolute value, and can be measured only
if very high experimental efficiencies and luminosities are achieved;
furthermore, the total cross section asymmetry is sizeably
contaminated by the hadronic process~\cite{SVdeltag}.
The situation improves when considering more exclusive
quantities; in ref.~\cite{FRdeltag} it was found that at moderate
$p_{\sss T}$ values the asymmetry for the point-like component can be
rather large, well above the minimum observable value (in this region,
the experimental efficiency is sizeable~\cite{Eichler96}). In
ref.~\cite{SVdeltag} it was argued that the hadronic component should
have a negligible impact in this case.

It can be concluded that heavy-quark photo- and electroproduction studies
at HERA are still in an early stage. Much more statistics will become
available in the near future, allowing more accurate tests of QCD. The
agreement of present data on charm total cross sections with the NLO
QCD calculations are satisfactory for choices of parameters consistent
with low-energy fixed-target results. 
Transverse momentum and
rapidity distributions have also been presented, and a general
agreement with the theory was found. Some discrepancies, as found for
example in the rapidity distributions, are not statistically
compelling as yet. With higher statistics and improved efficiencies,
measurements of the gluon density in the proton will greatly improve.

\section*{Acknowledgements}
I wish to thank the Organizing Committee for the warm hospitality in
Durham.  I also thank Stefano Frixione and Jenny Williams for
useful suggestions.

\end{document}